\documentclass[aps,pre,twocolumn,groupedaddress,showpacs]{revtex4}
\newif\ifpdf            
\ifx\pdfoutput\undefined \pdffalse \else \pdftrue \fi
\ifpdf 
\usepackage[pdftex]{graphicx}
\pdfcompresslevel9 \DeclareGraphicsExtensions{.jpg,.pdf,.png,.mps}
\else 
\usepackage[dvips]{graphicx}
\DeclareGraphicsExtensions{.eps,.ps} \fi
\usepackage{bm}

\begin{document}

\title{UV-Manipulation of Order and Macroscopic Shape
in Nematic Elastomers}

\author{P.M.~Hogan}
\author{A.R.~Tajbakhsh}
\author{E.M.~Terentjev}

\affiliation{Cavendish Laboratory, University of Cambridge,
Madingley Road, Cambridge, CB3 0HE, U.K. }

\date{\today}

\begin{abstract}A range of monodomain nematic liquid crystal
elastomers containing differing proportions of photo-isomerisable
mesogenic moieties, which turn from a rod-like to a kinked shape
upon ultraviolet (UV) irradiation, was studied. Depending on the
proportion and positional role of the photo-sensitive groups in
the crosslinked polymer network, different types and magnitudes of
response were found. The principle consequence of such
photo-isomerisation is the destabilisation of the nematic phase,
whose order parameter depends on temperature in a near-critical
fashion. Accordingly, the effect of UV-irradiation is dramatically
enhanced near the critical temperature, with the associated
reduction in the nematic order parameter manifesting as a change
in the macroscopic shape of the elastomer samples, producing a
large uniaxial contraction. Theoretical analysis of this
phenomenon gives a good quantitative agreement with experiment.
\end{abstract}

\maketitle

\section{Introduction}

The unique behaviour of liquid crystal elastomers derives from the
intimate relationship between the elastic nature of the polymer
network, and the ordering of their mesogenic, liquid crystalline
moieties. Many interesting and unusual properties of these
materials have been identified, and are summarised in recent
review articles \cite{fink89,ober93,brand98,emt99,WT96}.

The property that is most relevant to this paper is the response
of a liquid crystal elastomers macroscopic shape to temperature. A
permanently, uniaxially aligned monodomain nematic liquid crystal
elastomer network \cite{kupfer94} will exhibit a spontaneous
contraction along its director axis when heated towards its
nematic-isotropic phase transition temperature
\cite{assfalg99,wermter01}. This phenomenon is due to the coupling
between the average polymer chain anisotropy, and the nematic
order parameter Q(T) \cite{WT96}. The large variation in magnitude
of resulting spontaneous uniaxial deformation depends strongly on
the extent of this coupling between the elastic and the
liquid-crystal degrees of freedom in the material
\cite{static01,thermal01}.

Accordingly, a mechanical action can be produced in response to a
control signal which defines the underlying order parameter of the
elastomer. Substantial macroscopic movement and shape changes are
brought about by subtle variations in microscopic ordering. There
are many methods of achieving mechanical motion in response to a
control signal - or actuation - for many different applications
\cite{fleck97}, ranging from precise piezoelectric positioning
devices to artificial muscles \cite{degen97,ratna01}. Liquid
crystal actuators predominantly rely on thermal
\cite{wermter01,thermal01,ratna01} and, to a much lesser extent,
electromagnetic \cite{meyer97} control signals. Recently, however,
a less invasive method of manipulating the underlying nematic
order parameter in an elastomer network has been reported for the
first time \cite{pereira01}.

\begin{figure}
\centering \resizebox{0.35\textwidth}{!}{\includegraphics{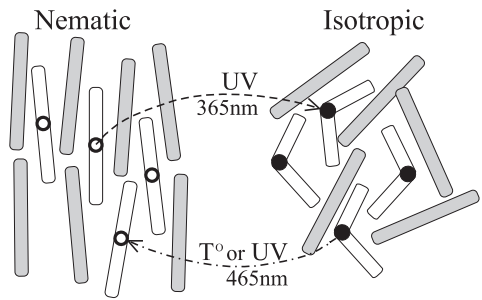}}
\caption{Schematic depiction of nematic-isotropic phase
transformation in a liquid crystal containing photo-isomerisable
mesogenic molecules, which turn from a rod-like {\it trans} to a
kinked {\it cis} conformation under the UV irradiation.}
\label{fig1}
\end{figure}

When exposed to ultraviolet (UV) radiation, liquid crystal
molecules containing photo-isomerisable groups, such as
azobenzene, experience a reduction in their nematic order and can
therefore be induced to undergo a phase transformation from the
nematic to the isotropic state \cite{ikeda87,ikeda99,legge92}.
This is due to the UV-induced {\it cis-trans} isomerisation of the
azo- (N=N) bond, whereby the photo-sensitive mesogenic molecules
change from a rod-like shape, to a strongly kinked shape upon
irradiation at a resonant wavelength of 365nm
\cite{pereira01,ikeda99,janossy98}. The rod-like shape serves to
stabilise the liquid crystal phase, whereas the kinked shape has
the opposite effect, acting as an impurity and destabilising the
nematic phase, reducing its order parameter \cite{ikeda99},
Fig.~\ref{fig1}. The dynamics of this process, and the reverse
{\it trans}$\rightarrow${\it cis} reaction occurring on heating or
irradiation at 465nm, are well studied in ordinary low-molar
weight liquid crystals \cite{ikeda99,kim95}.

With this method of non-invasive control over the degree of
nematic ordering, there is now a new mechanism for actuation in
liquid crystal elastomers. Such mechanical photo-induced actuation
in nematic elastomers was first reported by Finkelmann et al.
\cite{pereira01}. The degree of wide-spread, far-reaching interest
in this new photo-mechanical effect is highlighted by recent
popular articles in the New Scientist and The Financial Times
\cite{ft}, which appear even before any in-depth research had been
conducted, and suggest exotic applications such as surgical
micro-actuators.

In this article, a range of such photo-sensitive nematic liquid
crystal elastomers is studied. Before investigating their varied
response to ultraviolet light, their thermal responses are first
characterised in section~\ref{thall}. The differing UV responses
for the different sample compositions are presented in
section~\ref{uvall}. Section~\ref{azo18} then focuses on the
detailed quantitative analysis of the UV-mechanical response,
comparing the results with theoretical expressions derived in
section~\ref{theo}.

\section{Experimental}

\begin{figure*}
\centering \resizebox{0.7\textwidth}{!}{\includegraphics{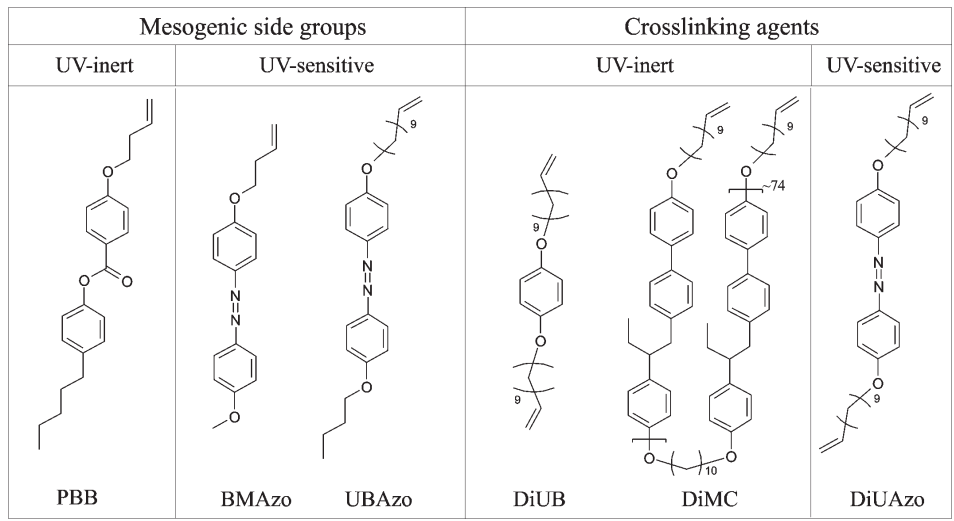}}
\caption{The mesogenic moieties used are abbreviated as follows:
PBB - 4-pentylphenyl-4'-(4-buteneoxy)benzoate; BMAzo -
[4-(4-buteneoxy)-4'-methyloxy]azobenzene; UBAzo -
[4-(11-undeceneoxy)-4'-butyloxy]azobenzene; DiUB -
di-1,4-(11-undeceneoxy)benzene; DiMC -
$\alpha$-{4-[1-(4'-{11-undeceneoxy}biphenyl)-2-phenyl]butyl}-$\omega$-(11-undeceneoxy)
poly-[1-(4-oxydecamethyleneoxy)-biphenyl-2-phenyl]butyl ($\sim$75
units long); DiUAzo - di-[4-(11-undeceneoxy)]azobenzene. }
\label{tab1}
\end{figure*}

All starting materials and resultant aligned, monodomain nematic
liquid crystal elastomers were prepared in the Cavendish
Laboratory. The procedure for forming the side-chain polysiloxanes
by the hydrosilation of the terminal vinyl group in the mesogenic
moiety with the Si-H bond of the polysiloxane chain, as well as
the two-step crosslinking technique with a uniaxial stress applied
after the first stage of crosslinking to produce and freeze the
monodomain nematic alignment, has been developed over the years by
Finkelmann et al. \cite{kupfer94,wermter01,greve01}. A number of
minor modifications were made to the procedure, which are
described in \cite{thermal01}. The azobenzene compounds were all
synthesised according to standard literature techniques
\cite{boden83,vogel,moreuv}. A summary of the different mesogenic
moieties used is given in Fig.~\ref{tab1}.

From the structure of the molecular moieties one can deduce that
all the side-group compounds are mesogenic rod-like units. PBB and
BMAzo have a four carbon atom spacer, which induces a parallel
orientation of the rods to the siloxane backbone resulting in a
prolate chain anisotropy in the nematic phase, with its principal
radii of gyration $R_\| > R_\bot$ . The third rod-like side group,
UBAzo, has a flexible spacer of eleven carbon atoms and is thus
only expected to be weakly coupled to the siloxane backbone. There
are three distinct types of di-vinyl crosslinking agents: the
flexible, non-mesogenic DiUB, the rod-like DiUAzo and the
main-chain nematic polymer DiMC, which contains a total of
approximately 75 rod-like monomer units between its reacting vinyl
groups. The polymer backbone onto which all these compounds were
grafted was polyhydromethylsiloxane (PHMS), containing
approximately 60 SiH groups per chain, obtained from ACROS
Chemicals.

\begin{center}
\begin{table*}
\caption{Compositions, and corresponding abbreviations, of the
different photo-sensitive nematic elastomers studied in this work.
The composition is presented in two forms: first, in mol\% of
corresponding moieties; second, the proportion of rod-like
mesogenic units in the overall composition. The rod-like side
groups are a combination of a ``standard'' nematic PBB and one of
two kinds of the azo-containing groups (cf. Fig.~\ref{tab1}),
specified in the second column of the table. The crosslinking
density in all cases is kept at a constant 10mol\%. \label{tab2} }
\begin{tabular*}{\textwidth}{|c|c|c|c|c|c|c|c|c|c|c|c|c|c|}
 &  & \multicolumn{6}{c}{Composition (mol\%)} &
                           \multicolumn{6}{c}{Composition (rod\%)} \\
Samples & Azo & \multicolumn{2}{c}{\underline{Side group}} &
\multicolumn{3}{c}{\underline{Crosslinking agent}}
& Total & \multicolumn{2}{c}{\underline{Side group}} &
\multicolumn{3}{c}{\underline{Crosslinking agent}} & Total \\
& moiety &  {\small PBB} & {\small Azo} & {\small DiUB} & {\small
DiUAzo} & {\small DiMC} & \{{\small Azo}\} &
{\small PBB} & {\small Azo} & {\small DiUB} & {\small DiUAzo} &
{\small DiMC} & \{{\small Azo}\} \\
 \hline
DiAzo100 & {\small BMAzo} & - & 90 & - & 10 & - & 100 & - & 90 & -
& 10 & - & 100 \\
Azo30 & {\small UBAzo} & 60 & 30 & 10 & - & - & 30 & 67 & 33 & - & - & - & 33 \\
Azo18 & {\small BMAzo} & 72 & 18 & 10 & - & - & 18 & 80 & 20 & - & - & - & 20 \\
Azo18MC & {\small BMAzo} & 72 & 18 & 9 & - & 1 & 18 & 44 & 11 & - & - & 46 & 11 \\
DiAzo9 & {\small BMAzo} & 92 & - & 1 & 9 & - & 9 & 91 & - & - & 9 & - & 9 \\
  \hline
\end{tabular*}
\end{table*}
\end{center}

The polymer networks were all crosslinked via the same
hydrosilation reaction, in the presence of a commercial platinum
catalyst COD, obtained from Wacker Chemie. The crosslinking
density was 10 mol\% (molar percent, relative to the total number
of SiH units) in all resultant elastomers.

Phase sequences were established on a Perkin Elmer Pyris 1
differential scanning calorimeter, which correlated with the
critical temperatures obtained by the thermal expansion
measurements (section~\ref{thall}). In all resulting elastomeric
materials a broad nematic liquid crystalline phase below a certain
critical temperature was observed, with optical microscopy and
mechanical testing confirming that this was indeed the nematic
phase. Some of the materials also possessed a second ordered phase
at low temperatures, exhibiting high rigidity and little response
to temperature or UV-irradiation, just above the glass transition.
This additional phase was not a focus of the present work - all
measurements were performed in the soft rubbery nematic phase of
the elastomer samples.

Elastomer mechanical expansion measurements were obtained by
freely suspending the samples in an active, glass-fronted
Gallenkamp oven and measuring their variation in length with a
travelling microscope. Cooling and heating rates were chosen such
that thermal hysteresis effects were negligible (around
0.5${}^{\rm o}$C/min, except for the sample containing DiMC, which
required a much slower rate).

Samples (of approximate dimensions $30\times 5 \times 0.4$mm) were
uniformly irradiated with a Merck VL-4.L UV lamp, providing a
narrow band at 365nm with an output of approximately 4W, at a
distance of approximately 15cm. The temperature was determined
with a thermocouple placed directly behind the samples.

\section{Comparison of Material Properties}  \label{comp}

A wide range of elastomer compositions and topologies (summarised
in the Table~\ref{tab2}) was investigated, in order to probe
different aspects of nematic elastomer response to UV-irradiation.
It was universally found that the main consequence of
UV-irradiation was the {\it cis-trans} isomerisation of the
azobenzene groups, with a very low probability of side reactions.
However, the differing compositions and topologies of the samples
resulted in a range of substantially different magnitudes of the
effect. One key factor to consider is the overall concentration of
isomerisable azobenzene groups, which varied between 9 and
100mol\% in our materials. Another crucial factor is the
topological role of the crosslinker (cf. Fig.~\ref{tab1}). The
small, flexible crosslinking polymer, DiUB is deemed to have a
minor effect on the overall mesogenic properties of the liquid
crystal elastomer. This is quite unlike the other two crosslinkers
- DiMC is a highly anisotropic main-chain nematic polymer
containing $\sim$75 rod-like monomer units, which can align itself
and fold into hairpin defects in the nematic phase. DiUAzo
contains a single, rod-like photo-isomerisable unit of its own;
its effect on the mechanical response of the elastomer network to
UV-irradiation was the subject of the pioneering study by
Finkelmann et al. \cite{pereira01}.

\subsection{Uniaxial thermal expansion}  \label{thall}

\begin{figure}
\centering \resizebox{0.49\textwidth}{!}{\includegraphics{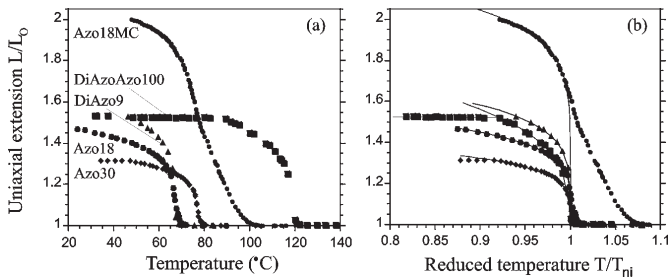}}
\caption{Uniaxial thermal expansion of nematic elastomers. Plot
(a) shows the raw data for the extension of the freely suspended
elastomers. With plot (b), the temperature for each sample is
scaled with the extrapolated nematic-isotropic transition
temperature $T_{\rm ni}$. Below this point, each data set is
fitted with a critical power law of the form $(\lambda-1) \propto
|T - T_{\rm ni}|^\xi$. The main-chain containing sample Azo18MC
shows a pronounced pre-transitional expansion (see text).}
\label{thermal}
\end{figure}

Before their response to ultraviolet light was investigated, the
spontaneous uniaxial thermal expansion of each of the samples was
first characterised (Fig.~\ref{thermal}). This determines the
extent of the coupling between the local nematic order, described
by the scalar order parameter $Q(T)$, and the average elastomer
network anisotropy, measured as the spontaneous uniaxial extension
$\lambda=L/L_0$, where $L$ is the current length of the sample,
and $L_0$ its length in the isotropic phase (the latter is always
shorter for an elastomer with a prolate backbone configuration, as
is the case for all our materials).

Empirically, one can relate the uniaxial thermal expansion of a
liquid crystal elastomer, $\lambda$, to the underlying nematic
order parameter $Q(T)$, which can be independently determined by
either wide-angle X-ray scattering, or from the optical
birefringence of the elastomer. It has been shown
\cite{static01,greve01} that the relationship between the
macroscopic shape and the nematic order of liquid crystal
elastomers can be very well described by a direct linear
proportionality, $\lambda = 1 + \alpha \, Q$. The constant
$\alpha$ is a sensitive function of the average network
anisotropy, varying from $\alpha \sim 0.08$ in weakly anisotropic
elastomers to $\alpha \sim 3$ or greater in materials containing
main-chain crosslinkers \cite{static01}. Such a relationship
between the spontaneous uniaxial extension and the underlying
nematic order is well-understood theoretically within the
molecular model of equilibrium nematic polymer networks
\cite{WT96}.

However, a conceptual problem resides in the values of the nematic
order parameter itself. The thermal expansion data sets for all
our materials, as well as the analogous studies by other groups,
e.g. \cite{greve01,ratna01}, were all found to follow $\lambda = 1
+ \alpha \, Q$ very faithfully, but with $Q$ varying in the manner
of a virtual second-order phase transition, $Q \propto |T - T_{\rm
ni}|^\xi$ \cite{chaikin}. In our present materials the critical
exponent of the apparent critical behaviour, $\xi$, takes values
between 0.14 and 0.31 (Table~\ref{tab3}). There are several
possible reasons for this observation of an apparent critical
behaviour in a system that, by symmetry arguments, should undergo
a discontinuous first order transition. One likely explanation is
that the kinetic response of the rubber-elastic network is such
that it is unable to perform the discontinuous jump required by
the changing underlying nematic order parameter. Such an
explanation, although appealingly simple, has several weaknesses,
the main one being that practically no dependence on
cooling/heating rate is observed, which should be expected if the
kinetic retardation is at work.

Another possibility is that in forming the monodomain nematic
materials, an effective aligning molecular field is frozen in by
crosslinking. This makes the first-order nematic-isotropic
transition supercritical and, essentially, eliminates the true
isotropic phase altogether:  even at high temperature there would
still be a small residual order remaining. This could still be the
case, even though in our materials we could not experimentally
detect any residual order above $T_{\rm ni}$ and observe a clear
and well-defined heat signature of the transition on DSC.

A perhaps more intriguing possibility is the presence of quenched
sources of random disorder in the nematic field (the network
crosslinking points), which genuinely transforms the weak first
order transition into a continuous one, an effect already
discussed in the context of spin glasses \cite{cardy99}. The
present paper does not attempt to address this issue. We focus
instead on a different physical property of a photo-sensitive
nematic elastomer and simply use this experimentally observed,
continuous ``critical'' variation of $Q(T)$ as the foundation for
describing the UV-induced response of such an elastomer.

\begin{table}
\caption{Experimentally determined transition points, $T_{\rm ni}$
(used as a parameter in fitting by the power-law), and the
apparent critical exponents, $\xi$, for all materials studied. The
samples are ranked by their overall azo-content. \label{tab3} }
\begin{center}
\begin{tabular}{|c|c|c|}
Sample & $T_{\rm ni}^{(0)}$ (${}^{\rm o}$C) & $\xi$ \\
 \hline
DiAzo100 & 120.3 & 0.31 \\
Azo30 & 76.8 & 0.21 \\
Azo18 & 67 & 0.19 \\
Azo18MC & 75.2 & 0.14 \\
DiAzo9 & 66.5 & 0.20 \\
  \hline
\end{tabular}
\end{center}
\end{table}

It seems that a consistent value of the ``critical exponent'' is
obtained, $\xi \sim 0.2$. The anomalously small value of $\xi$ for
Azo18MC can be attributed to the significant difference in this
sample's response time from all the others: kinetic effects are
markedly relevant and the long DiMC chains have been shown to slow
down, and even completely freeze their dynamics, when in the
hairpin regime of their nematic conformation \cite{elias99}. This
would result in the non-equilibrated thermal expansion of the
sample lagging the actual, equilibrated order parameter. This
sample also shows a very significant pre-transitional effect, with
the residual nematic order and apparent sample extension
persisting far above the critical temperature. This phenomenon has
been examined in a separate study \cite{thermal01} and is due to a
composite nature of this material in which the side-chain and
main-chain polymer chains are, crudely, in equal proportion.
Another point of interest is the presence of highly ordered phases
in DiAzo100 and Azo30 at lower temperatures, where the dependence
of $Q$, and hence $\lambda$, on temperature deviates from the
attributed nematic behaviour, and becomes essentially constant.
However, this will not be dwelt upon, as this work focuses
exclusively on the nematic phase and the nematic-isotropic phase
transition.

The observed dependence of $Q$ and $\lambda$ on temperature is
very important in analysing and modelling the response of the
liquid crystal elastomers to UV-irradiation. In doing so, one
sample (Azo18), which proved to have the most pronounced response
to UV-irradiation, is studied in greater detail in
Section~\ref{azo18}.

\subsection{Response to UV-irradiation}  \label{uvall}

In order to compare different compositions and crosslinking
topologies of the nematic elastomers in their response to
UV-irradiation, all samples were equilibrated before irradiation
at approximately $10^{\rm o}$C below their critical temperature
$T_{\rm ni}$, determined by the thermal analysis in section
\ref{thall}. Such proximity to $T_{\rm ni}$ also aims to maximise
the variation of $\lambda$, whilst avoiding the singularity at the
phase transition.

\begin{figure}
\centering \resizebox{0.4\textwidth}{!}{\includegraphics{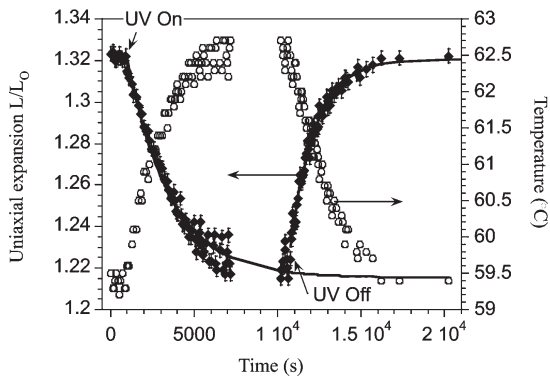}}
\caption{The response of Azo18 to UV irradiation. Both the
relative length of the sample (uniaxial strain $\lambda$, measured
with respect to the isotropic state: filled symbols) and the
measured temperature (open circles) are shown. The $\pm 0.1^{\rm
o}$C errors in temperature have been omitted for clarity. Solid
lines are fits of both the {\it on} and {\it off} branches by a
simple exponential, which at this point should be regarded as only
a guide to the eye. } \label{uv1}
\end{figure}

During irradiation, the sample temperature was, without exception,
found to increase in an approximately exponential manner,
mirroring the behaviour of the spontaneous uniaxial extension,
$\lambda$. Similarly, as the elastomers relaxed after the
radiation was switched off, the measured temperature again
mirrored $\lambda$, decreasing back to its pre-irradiated value.
In both cases, this exponential variation in temperature was found
to be modulated by a small-amplitude oscillatory behaviour.
Possible explanations for this thermal phenomenon involve a
trivial heating effect from the UV lamp interfering with the
active temperature control of the oven. However, such an
additional heating power should not result in the overall rise of
saturated temperature value, as seen in Fig.~\ref{uv1}: the oven
would be expected to reduce its own heating and bring the
temperature back to its set value. Also, it would be an
inconceivable coincidence if the characteristic rate of the oven
heating system are exactly the same as that of all our different
materials.

By closely examining the Fig.~\ref{uv1} it becomes clear that we
observe the effect of spontaneous sample heating, occurring due to
the exothermic nature of isomerisation of azobenzene groups. In a
following section we shall describe the population dynamics
between the {\it trans} and {\it cis} N=N isomers. The relative
proportion of these states is controlled by the rates of forward
and reverse reactions, which reach the balance in the saturated
state (after a long time of irradiation). If one assumes that the
heat release on the return to the {\it trans} ground state is
proportional to the rate of change of {\it cis} concentration,
$\dot{n}_{cis}$, then the observation in Fig.~\ref{uv1} becomes
sensible. Initially, all azobenzene groups are in their {\it
trans} state and the sample is thermally equilibrated with the
oven (at 59.5${}^{\rm o}$C in this case). On irradiation, the
increasing number of {\it cis} isomers is created, and also
accompanied by an ever increasing relaxation flux back to the {\it
trans} state. This heats the sample, at the rate exactly matching
the rate of isomerisation itself. Finally, a saturation state is
reached, when the UV-stimulated {\it trans}$\rightarrow${\it cis}
flux is equal to the relaxation flux: at this stage the heat
release in the sample is constant and maximal. On switching the
UV-light off the exothermic {\it cis}$\rightarrow${\it trans}
relaxation continues to heat the sample, but with a decreasing
power as the overall concentration ${n}_{cis}$ decreases.

All the samples responded to UV-irradiation in a manner similar to
Azo18, illustrated in Fig.~\ref{uv1}, differing only in the
magnitude and speed of their response. The resultant contraction
upon irradiation for each sample, relative to the expected
contraction due to the aforementioned increase in temperature, are
summarised in Table~\ref{tab4}. As can be seen, the proportion of
UV-isomerisable azobenzene rods in the sample is by no means a
direct indication of its response to UV-irradiation, interpreted
as the maximal sample extension achieved on irradiation.

\begin{table}
\caption{Sample contractions upon UV-irradiation, relative to the
expected thermal contraction due to the accompanied temperature
increase above the starting point (last column). The increase of
this ration above unity is a genuine measure of the effect of UV
isomerisation. \label{tab4} }
\begin{center}
\begin{tabular}{|c|c|c|}
Sample & $\Delta \lambda/ \Delta \lambda(T)$ &   $T_{\rm exp}$ (${}^{\rm o}$C)\\
 \hline
DiAzo100 & 1.21 & 110.5 \\
Azo30 & 1.49 & 66.9 \\
Azo18 & 2.82 & 57 \\
Azo18MC & 1.03 & 65.0 \\
DiAzo9 & 1.63 & 56.8 \\
  \hline
\end{tabular}
\end{center}
\end{table}

Azo18 gave the largest response to radiation, but the very similar
Azo30, with its higher proportion of azo-rods, gave a
significantly smaller response. This must be due to the position
of the azo-group within the corresponding mesogenic moiety - with
Azo30, it is only weakly coupled to the polymer backbone by a long
spacer, in contrast to Azo18, where a strong parallel alignment
between the rod and the backbone is induced. This could result in
the kinked {\it cis}-form of the moiety having less of a
destabilising effect on the backbone anisotropy, and so the
elastomer response to UV-irradiation is curtailed.

DiAzo9 is very similar to the material studied by Finkelmann et
al. \cite{pereira01}. Despite containing less than a third of the
proportion of azobenzene rods found in Azo30, its response to
UV-irradiation is comparable in magnitude, which should be
attributed to the crucial position of the photo-isomerisable
group, in the centre of the predominant, short crosslinker.

Azo18MC might have been expected to have a very large response
compared to the other materials due to its large backbone
anisotropy and large coupling constant to the underlying order
parameter, illustrated by its large thermal expansion amplitude.
However, this was definitely not the case for UV irradiation. One
may speculate the cause may be not only that the actual proportion
of azobenzene rods was quite low for this sample (much of the
volume is occupied by the main-chain polymer crosslinker),
resulting in a smaller absolute change in the order parameter, but
more importantly, that large DiMC chains can essentially freeze
the elastomer dynamics in response to changes in $Q$
\cite{elias99}, quenching its manifestation through $\lambda$.

All the mesogenic groups and crosslinkers in DiAzo100 were
photo-sensitive, and so one might have expected an exceptional
mechanical response to UV-irradiation. However, the critical
temperature, $T_{\rm ni}$, of this sample was much higher than
that of the other materials, and to maintain consistency between
experiments on different samples this required an initial
equilibrium temperature of approximately $110^{\rm o}$C. At such a
temperature the thermally-activated relaxation taking the
azobenzene groups from their metastable kinked {\it cis}-shape
back to the rod-like {\it trans}-shape would be much accelerated.
As a result, one is perhaps unable to isomerise a significant
proportion of azobenzene rods at any given moment of time. A lower
equilibrium population of {\it cis}-isomers would then result in a
smaller overall change in $Q$, and hence in $\lambda$.

The solid curves in Fig.~\ref{uv1} are simple exponential fits to
the data, for both the UV-on and the UV-off states. Apparently, it
is a reasonably good fit. However, the parameters are rather
arbitrary and appear to be quite unrelated to the corresponding
spontaneous thermal expansion curves. It will shortly be seen that
a different, universal theoretical dependence on temperature and
irradiation time describes the data with a much greater degree of
fidelity. Accordingly, the solid lines in Fig.~\ref{uv1} serve
only as a rough guide to the eye.

\section{Manipulation of Nematic Order} \label{azo18}

\subsection{Population dynamics} \label{dyna}

\begin{figure}
\centering \resizebox{0.18\textwidth}{!}{\includegraphics{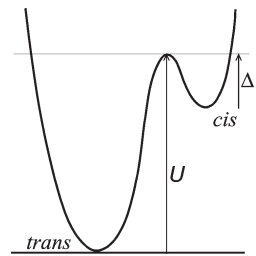}}
\caption{Schematic depiction of the potential energy profile of a
photo-sensitive moiety verses configuration, showing the relative
stabilities of the {\it cis} and {\it trans} isomers.}
\label{well}
\end{figure}

The internal energy of rod-like {\it trans}, and the kinked {\it
cis} mesogenic moieties is adequately described as a two-level
potential well, with the former configuration energetically more
stable than the latter (Fig.~\ref{well}). In the experiments
conducted, there are three possible ways of moving between states,
each characterised by a separate time constant:\\
 \noindent $\bullet $   \
Thermal excitations up, from {\it trans} to {\it cis}, with the
characteristic time given by $\tau_{tc} = \tau_0 e^{U/kT}$,
with $\tau_0$ the bare attempt rate.\\
 \noindent $\bullet $   \
Thermally driven relaxation down, from {\it cis} to {\it trans},
with the time constant $\tau_{ct} = \tau_0 e^{\Delta/kT}$.\\
 \noindent $\bullet $   \
§ UV-isomerisation, occurring with the rate $1/\tau_{\rm uv}
\equiv \eta \propto$ radiation intensity.

One expects that the barrier height $U$ is much greater than the
metastable state depth $\Delta$, and hence one can neglect the
rate of spontaneous, thermally driven {\it trans} to {\it cis}
transitions, although certainly not its reverse.

Let us assign the total (fixed) number density of
photo-isomerisable rod-like groups in a given elastomer as $n_0 =
n_{trans}+n_{cis}$, with the only independent variable defined as
$n \equiv n_{trans}$, so that $n_{cis}=n_0-n$. Thus, the rate of
change in the population of {\it trans} isomers can simply be
written as;
 \begin{equation}
\frac{ \partial}{\partial t} n(t,T) = -\eta \, n -
\frac{1}{\tau_{tc}}n + \frac{1}{\tau_{ct}}(n_0-n) \label{eq1}
 \end{equation}
In the above approximation, the relaxation time $\tau_{tc}$ is
assumed to be very large, and can therefore be neglected. The
appropriate initial condition is then $n_{t=0}=n_0$ (that is, all
azobenzene groups are in their rod-like {\it trans} conformation)
and the dynamic solution to Eq.~\ref{eq1} is simply;
 \begin{eqnarray}
n(t,T) &=& \frac{n_0}{1+\tau_{ct}\eta} \left[ 1+ \tau_{ct}\eta \,
\exp \left(-t/\tau_{\rm eff} \right) \right] \label{eq2} \\
& {\rm with} & \tau_{\rm eff}= \left( \eta + 1/\tau_{ct}
\right)^{-1} \nonumber
 \end{eqnarray}
Hence, one can see that the increasing concentration of kinked
{\it cis}-isomers, which act as impurities for the nematic
ordering, will eventually reach a plateau due to the opposing
thermal back-reaction, itself characterised by the rate
$1/\tau_{ct}$. Both the rate and the final saturation value of the
isomer density are crucially dependent on the irradiation rate
$\eta$; this could be different for different materials, sample
shapes and experimental conditions as it is affected, among other
factors, by the radiation absorption in the medium.

The relaxation process, after switching the radiation off (that
is, setting $\eta=0$), is the thermally driven flux of {\it
cis}-isomers into the {\it trans} ground state over the barrier
$\Delta$. Assuming a fully saturated irradiated state, with
$n_{cis}=n_0-n_{\infty}= \tau_{ct}\eta/(1+\tau_{ct}\eta)$ at $t=0$
(and still neglecting the spontaneous thermal {\it
trans}$\rightarrow${\it cis} process), results in the relaxation
law
 \begin{eqnarray}
 n(t,T) &=& n_0 \left[1- \frac{\tau_{ct}\eta}{1+\tau_{ct}\eta}
\exp \left(-t/\tau_{ct} \right) \right] . \label{relax}
 \end{eqnarray}
Naturally, the relaxation time scale is purely $\tau_{ct}$, a
property of an individual azobenzene group, which should remain at
least approximately universal between all different samples under
study.

\subsection{Macroscopic shape and nematic order}  \label{theo}

The influence of impurities on the nematic order is well studied
\cite{degen}. One of the simplest ways of representing their
effect is via the Landau-de Gennes mean-field formulation, which
describes the local free energy density as a power series in the
magnitude of the scalar order parameter, $Q$. Such an approach is
commonly used in the thermodynamics of second order phase
transitions, as well as in application to liquid crystals.
Although it is much less suitable for the nematic-isotropic
transition in liquid crystals, which should be first-order by
symmetry, many useful qualitative results are obtained within the
Landau-de Gennes formalism.

\begin{figure}
\centering \resizebox{0.35\textwidth}{!}{\includegraphics{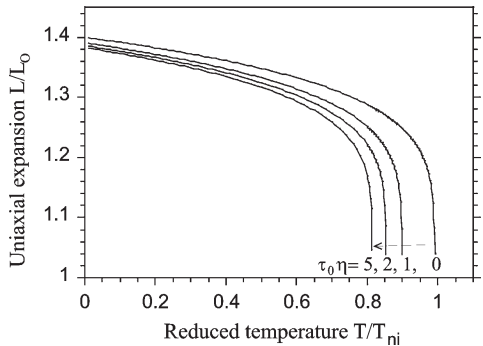}}
\caption{The theoretically calculated effect of increasing the
intensity of UV-irradiation (measured by the effective
non-dimensional rate $\tau_0\eta$) on the uniaxial thermal
extension of a photo-sensitive liquid crystal elastomer, according
to Eq.~\ref{eq5}.} \label{theory}
\end{figure}

However, in view of many limitations of such a phenomenological
mean-field theory, we adopt a different approach. In
section~\ref{thall}, the actual form of the dependence of the
order parameter on temperature was established experimentally as
being a virtual, continuous transition with the apparent critical
exponent $\xi$ quite far from the mean-field value predicted by
the Landau-de Gennes approach (and certainly appears a continuous
transition). Therefore, we simply take the empirical dependence,
$Q \propto |T - T_{\rm ni}|^\xi$, and modify it by substituting in
the impurity-shifted critical temperature, $T_{\rm ni}=T_{\rm
ni}[n(t,T)]$.

The effect of impurities on the critical temperature is, in the
first approximation, a linear function of the population of {\it
cis}-isomers. The size of this population, $n_{cis}$, defined
above as $n_0 - n(t)$, is itself a function of time;
 \begin{equation}
 T_{\rm ni}(n)=T_{\rm c}-\beta [n_0-n(t)]  ,  \label{eq4}
 \end{equation}
where $T_{\rm c} \equiv T_{\rm ni}^{(0)}$ is the transition
temperature in the ``pure'' material containing only the {\it
trans} rod-like units (given in the Table~\ref{tab3} for our
materials). Combining this with the known dependence of the
spontaneous uniaxial sample extension, $\lambda$, on $Q$, one
obtains the following relation to describe the observed mechanical
response to UV-irradiation
 \begin{eqnarray}
&&\lambda_{\rm uv}(t,T) = \label{eq5} \\
&&= 1+\alpha \left[ T_{\rm c}- \beta \, n_0 \left( 1-
 \frac{\tau_{ct}\eta}{1+\tau_{ct}\eta} (1+e^{-t/\tau_{\rm eff}})
\right) - T \right]^\xi \nonumber
 \end{eqnarray}
Plotting $\lambda$ against temperature for different values of the
intensity parameter $\eta$, Fig.~\ref{theory}, one sees that the
expected consequence of UV-irradiation is simply to lower the
critical temperature of the nematic-isotropic phase transition.

The relaxation of sample shape after switching the radiation off
is determined by the rate of recovery of the initial population of
{\it trans} isomers, according to Eq.~(\ref{relax});
 \begin{eqnarray}
\lambda_{\rm off}(t,T) = 1+\alpha \left[ T_{\rm c}- \beta \, n_0
\frac{\tau_{ct}\eta}{1+\tau_{ct}\eta} e^{-t/\tau_{ct}} - T
\right]^\xi , \label{relax2}
 \end{eqnarray}
where, as in the Eq.~(\ref{relax}), we have assumed that a fully
saturated state was achieved on preceding irradiation.

One important parameter has to be used in the further analysis:
the relaxation time scale of the {\it cis}$\rightarrow${\it trans}
process. From the data in Fig.~\ref{uv1} for the Azo18 at $\sim
59^{\rm o}$C it is estimated as $\tau_{ct} \approx 1479$~s,
assuming that the dynamics is dominated by the rate of
isomerisation (and not, for instance, the viscoelastic relaxation
of a rubbery network). In the following analysis of experimental
data, the effect of (Arrhenius) temperature variation of
$\tau_{ct}$ within a given data set are, in the first
approximation, neglected. This is justified by the relatively
narrow range of absolute temperatures over which a given data set
varies. However, across data sets, $\tau_{ct}$ is shown to vary
significantly with temperature.

\subsection{Nematic destabilisation by UV radiation} \label{trio}

\begin{figure}
\centering \resizebox{0.4\textwidth}{!}{\includegraphics{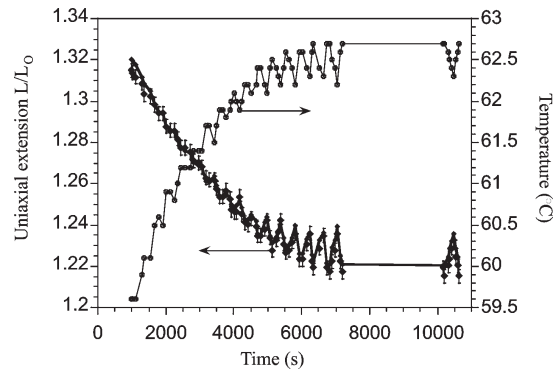}}
\caption{The response of Azo18 to UV-irradiation, and the
corresponding theoretical fit by Eq.~(\ref{eq5}), solid line. Note
that even the minor intrinsic temperature oscillations are, in
fact, directly manifested in $\lambda(t,T)$, and are almost
perfectly described by the theoretical model (see text).}
\label{uvfit}
\end{figure}

Applying the Eq.~(\ref{eq5}) to the data in Fig.~\ref{uv1}, one
obtains a remarkably improved fit (Fig.~\ref{uvfit}). In analysing
the data we take the fixed values for the coefficient
$\alpha=0.22$ and the exponent $\xi=0.19$, as determined by the
thermal expansion measurements of section~\ref{thall} and the
relaxation time $\tau_{ct}=1479$~s, as estimated from a crude
exponential fit of Fig.~\ref{uv1}. The values of current
temperature $T$ are taken from the corresponding experimental data
points. The free parameters are, thus, only the radiation rate
$\eta \Rightarrow 2.2 \cdot 10^{-4}\hbox{s}^{-1}$ and the product
$\beta \, n_0 \Rightarrow 11.8$, which is a measure of the overall
concentration, and the efficiency, of azobenzene groups.

The result of such a fit for $\lambda(t,T)$ is very satisfying.
The initial quasi-linear contraction and its subsequent rapid
flattening and saturation are fully described, unlike the rough
exponential fit in Fig.~\ref{uv1}. More importantly, we now
discover that what looked like noise in both the temperature and
the strain $\lambda(t)$ in the initial graph, Fig.~\ref{uv1}, is
in fact a consistent small-amplitude temperature oscillation
(probably due to the active oven control operation), which is also
fully reproduced in the mechanical shape of the sample. This is a
rather stringent test of the theoretical model and the parameters
used to fit the experimental data.

\begin{figure*}
 \centering \resizebox{0.96\textwidth}{!}{\includegraphics{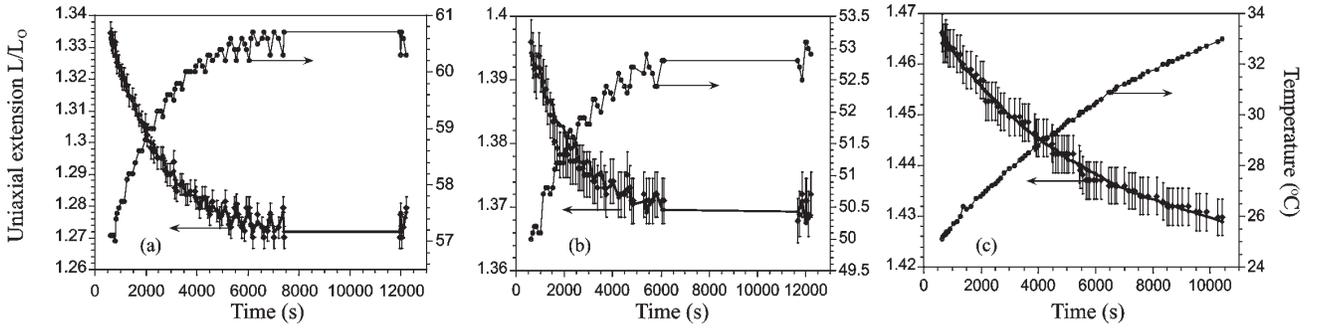}}
\caption{The response of Azo18 to UV-irradiation at different
starting temperatures, $\sim 57, \, 50$ and $25^{\rm o}$C for
plots (a), (b) and (c), respectively. The corresponding
description of $\lambda(t,T)$ with the theoretical model,
Eq.~(\ref{eq5}), is shown by the solid lines. The relative error
increases for lower temperatures due to the smaller overall
magnitude of the UV-response. } \label{uvfit3}
\end{figure*}

\begin{figure}
\centering \resizebox{0.49\textwidth}{!}{\includegraphics{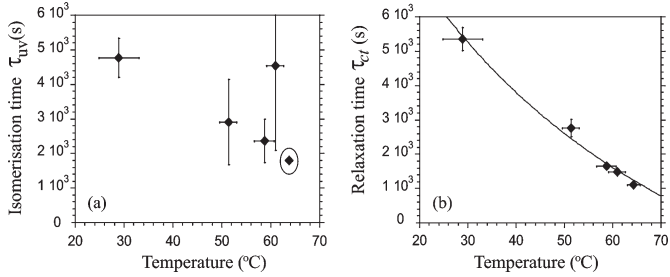}}
\caption{(a) The irradiation time constants, $1/\eta$, at
different temperatures, are all of the same order of magnitude and
show a constant or slowly decreasing dependence on temperature.
The circled point, near $T_{\rm ni}$, has an ambiguously high
error associated with it (see section~\ref{crit}).  (b) Relaxation
times $\tau_{ct}$ gleaned from a simple exponential fit of the
relaxation data. The activation temperature dependence allows a
crude estimate of the isomerisation energy barrier $\Delta \sim 4
\cdot 10^{-20}\hbox{J} \sim 10 kT$. } \label{times}
\end{figure}

All the other data sets for Azo18 at different temperatures are
fitted equally well, as shown in Fig.~\ref{uvfit3}. Once again, we
fix the value of strain-order coupling coefficient $\alpha=0.22$
and the exponent $\xi=0.19$, as determined by the thermal
expansion measurement for this sample, and the relaxation time
$\tau_{ct}=1479$~s. On physical grounds, we also insist that the
coefficient $\beta \, n_0$ remains constant ($\sim 12$) between
these data sets, since it is determined by the intrinsic
properties of the nematic order and the given concentration of
azobenzene groups. With these stringent constraints, one obtains
the values of the \underline{only} remaining free parameter, the
effective rate of induced isomerisation. The inverse of $\eta$,
the characteristic time for the photo-induced {\it
trans}$\rightarrow${\it cis} isomerisation, $\tau_{\rm
uv}=1/\eta$, is plotted in Fig.~\ref{times}(a) and has an average
value $\tau_{\rm uv} \sim 3 \cdot 10^3$s. Although there is a
certain spread in these values, we find the outcome of this
analysis quite satisfactory given the number of constraints
imposed on the fitting, certain error in the data and, in spite of
that, a clear evidence that not only the overall trend but also
the small oscillations of temperature are well reproduced by the
Eq.~(\ref{eq5}).

The values of the characteristic time for the reverse thermal {\it
cis}$\rightarrow${\it trans} relaxation, $\tau_{ct}$, are
consistent with those reported in the literature for azobenzene
photochemistry \cite{moreuv}, as well as within themselves,
Fig.~\ref{times}(b). The crude estimate of the {\it
cis}$\rightarrow${\it trans} energy barrier from the Arrhenius fit
of the relaxation time returns a reasonable value $\Delta \sim 4
\cdot 10^{-20}\hbox{J} \sim 10 kT$. This is in contrast with the
isomerisation time $\tau_{\rm uv}$, which was found to be a
constant, or very slowly decreasing function of temperature.
However, the errors in determining $\eta$ were large, and there is
much room for further studying its dependence on temperature.

\subsection{The vicinity of critical point} \label{crit}

Since the effect of UV radiation on a nematic rubber containing
isomerisable azobenzene groups is clearly identified as an
effective shift of critical temperature of the underlying
nematic-isotropic phase transition, the question arises - what
would be the response when the material is irradiated at a
temperature close to the ``bare'' critical point $T_{\rm c}$? In
this case, the increasing concentration of {\it cis}-impurities
should take the system past the transition line into the isotropic
phase. Continuing the study of Azo18, we now subject it to
UV-radiation at a temperature $T_{\rm exp}\approx 63^{\rm o}$C,
only 4${}^{\rm o}$ below $T_{\rm c}$.
\begin{figure}
\centering \resizebox{0.4\textwidth}{!}{\includegraphics{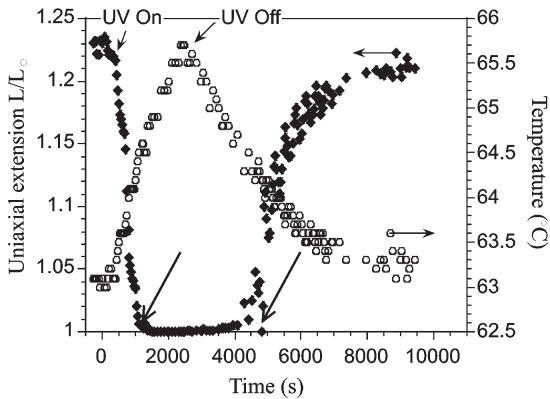}}
\caption{The response of Azo18 to UV-irradiation in the vicinity
of critical point. Arrows indicate the moments of time when the
material first becomes fully isotropic on irradiation, and then
re-enters the nematic phase on relaxation (the associated measured
temperature of $\sim 64^{\rm o}$C is clearly insufficient to
account for this transition). } \label{critplot}
\end{figure}

Fig.~\ref{critplot} indicates that the mechanical response of the
sample on irradiation, $\lambda(t)$, starts exactly in the same
way as at all other temperatures. However, the initial steep drop
in $\lambda$ (and in the underlying nematic order $Q$) rapidly
reaches the point of $\lambda=1$ (or $Q=0$); the material becomes
fully isotropic and no longer has any macroscopic response to the
further increase of the population of {\it cis}-isomers. The
irradiation continues for some time, as illustrated by the
continuing rise in temperature, which nevertheless has never
reached the ``bare'' $T_{\rm c}$. Then the UV-light is switched
off and the temperature begins to fall, in the same way as all
previous measurements have shown. However, nematic order is not
established for some time: clearly the concentration of {\it
cis}-impurities has to decrease sufficiently to allow the nematic
phase to appear at a given temperature.

Accordingly, there are two simultaneous thermal effects in the
material. One is the rise (in the UV-{\it on} state), or fall (in
the {\it off} state), of the actual temperature of the sample. The
other is the underlying shift of the critical temperature of the
nematic phase transition, described by the Eq.~(\ref{eq4}) with
the time-dependent concentration $n(t)$ given by Eqs.~(\ref{eq2})
and (\ref{relax}), respectively (in the latter case of thermal
{\it cis}$\rightarrow${\it trans} relaxation, Eqs.~(\ref{relax})
and (\ref{relax2}) assume that the state of full saturation has
been reached on irradiation, otherwise a different initial
condition has to be used, altering the prefactor to the
exponential). It is straightforward to apply the corresponding
analysis and determine the time when, in the UV-{\it on} state,
the decreasing critical temperature $T_{\rm ni}= T_{\rm c}-\beta
[n_0-n(t)]$ becomes equal to the increasing sample temperature.
Equally, in the UV-{\it off} state, we find the time needed for
the increasing $T_{\rm ni}$ to intersect the decreasing actual
temperature $T$.

Applying this analysis to Azo18 in Fig.~\ref{critplot}, one
obtains that the nematic order disappears at $t\approx 1100$~s and
subsequently re-enters the nematic phase, is $t\approx 4850$~s.
Both times are marked with arrows in Fig.~\ref{critplot}: clearly
the agreement with the observed change in the sample length,
$\lambda=L/L_0$, is excellent. This is another confirmation of the
theoretical model, which assumes simple thermal activation laws
for characteristic times, the linear dependence of critical
temperature on the concentration of {\it cis}-isomers and, much
less trivially, the empirical critical dependence of the nematic
order parameter $Q \propto |T-T^*|^\xi$.

\section{Conclusions}

The derivation of Eq.~(\ref{eq5}) in section~\ref{theo},
cumulating in the analysis of relaxation data sets in
Figs.~\ref{uvfit}-\ref{uvfit3}, makes it clear that, once the
number of kinked moieties reaches a steady-state population, the
resultant effect of UV-irradiation is equivalent to simply
reducing the critical temperature of the nematic-isotropic phase
transition by an amount
$$ \Delta T = \beta [n_0-n_{\infty}(T)] = \beta \, n_0
\frac{\tau_{ct}\eta}{1+\tau_{ct}\eta},$$ which is dependent on the
intensity of the ultraviolet light used (through $\eta$).
\begin{figure}
\centering \resizebox{0.35\textwidth}{!}{\includegraphics{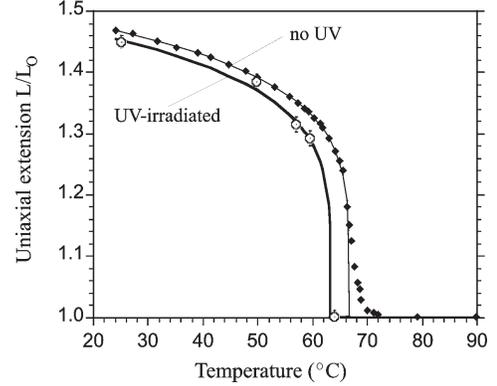}}
\caption{The asymptotic values of $\lambda(t,T)$, for Azo18
irradiated for a long time at several fixed temperature, after
correcting for the temperature increase associated with
irradiation, $\beta[n_0-n]$, which correspond well to the idea of
a shifted critical temperature $T^*$, as illustrated theoretically
in Fig.~\ref{theory}} \label{order}
\end{figure}
Hence, a slightly modified form of $\lambda=1+\alpha \, Q$, making
only the replacement $T_{\rm ni} \rightarrow \, T^*$, should fully
describe the final, steady-state length of the sample,
$\lambda_{\rm uv}(t\rightarrow \infty)$, as a function of
temperature. All the other parameters in Eq.~(\ref{eq5}), in
particular the exponent $\xi$, should remain unchanged. Indeed,
one finds that this idea does describe $\lambda_{\rm
uv}(t\rightarrow \infty)$ very well, as shown in Fig.~\ref{order}.
The corresponding value of shift $\Delta T$, which best fits the
data in Fig.~\ref{order}, is found to be $(3.3 \pm 0.7)^{\rm o}$C
which agrees well with the average asymptotic value of
$\beta[n_0-n_{\infty}]$ gleaned from extrapolating the data sets
in Figs.~\ref{uvfit}-\ref{uvfit3}, of $(3.5 \pm 1.9)^{\rm o}$C.

This work, has studied a wide range of photo-sensitive materials,
based on azobenzene molecular moieties, whose differing
compositions and topologies are reflected in their varied
mechanical response to irradiation with ultraviolet light. The
fundamental details of this response were investigated, in
particular its kinetics and the variation of the critical
temperature near the nematic-isotropic phase transition. The
obtained experimental results are very well described by a
theoretical model assuming a simple linear shift in the critical
temperature with the concentration of kinked {\it cis} groups,
which results in a rather non-trivial model for the dependence of
the underlying nematic order parameter $Q$ and macroscopic
uniaxial mechanical extension $\lambda$ on irradiation time and
temperature. Basic material parameters and constants were also
determined, which will allow subsequent investigations to further
probe the details of this fascinating phenomenon. With such a
novel, non-invasive method of control over mechanical actuation,
there is a great potential for application, but still very much is
left unknown.

\begin{acknowledgments}
We are grateful to Wacker Chemie for donating the platinum
catalysts for the chemical preparations, and to H. Finkelmann and
M. Warner for a number of useful discussions and advice. This work
has been supported by EPSRC UK.
\end{acknowledgments}

\end{document}